\documentclass[prb,superscriptaddress,longbibliography,twocolumn]{revtex4}
\usepackage{graphicx}
\usepackage{bm}
\usepackage{color}
\usepackage{url}
\usepackage{tikz}
\usetikzlibrary{calc}
\usepackage{pgfplots}

\newcommand{\bea}{\begin{eqnarray}}
\newcommand{\eea}{\end{eqnarray}}
\newcommand{\nn}{\langle ij \rangle}

\newcommand{\bS}{\mathbf{S}}

\newcommand{\br}{\mathbf{r}}
\newcommand{\hatz}{\hat{z}}

\newcommand{\tI}{\text{i}}
\newcommand{\te}{\text{e}}
\newcommand{\td}{\text{d}}
\newcommand{\mcE}{\mathcal{E}}

\newcommand{\be}{\begin{equation}}
\newcommand{\ee}{\end{equation}}
\newcommand{\bk}{{{\bf{k}}}}

\newcommand{\bQ}{{{\bf{Q}}}}

\newcommand{\beal}{\begin{align}}
\newcommand{\eeal}{\end{align}}
\newcommand{\ra}{\rangle}
\newcommand{\la}{\langle}

\newcommand{\dg}{{\dagger}}
\newcommand{\pdg}{{\phantom\dagger}}

\renewcommand{\bm}{\mathbf{m}}
\newcommand{\beas}{\begin{eqnarray*}}
\newcommand{\eeas}{\end{eqnarray*}}
\begin{document}

\title{Tunable skyrmion crystals and topological quantum oscillations in magnetic metals}
\author{Sopheak Sorn}
\affiliation{Department of Physics, University of Toronto, Toronto, Ontario, Canada M5S 1A7.}
\author{Stefan Divic}
\affiliation{Department of Physics, University of Toronto, Toronto, Ontario, Canada M5S 1A7.}
\author{Arun Paramekanti}
\email{arunp@physics.utoronto.ca}
\affiliation{Department of Physics, University of Toronto, Toronto, Ontario, Canada M5S 1A7.}

\date{\today}

\begin{abstract}
Skyrmions are spatially localized magnetic swirls which carry a nonzero integer topological charge. We study crystals of skyrmions in a
two-dimensional ferromagnet model with chiral interactions induced by the presence of broken inversion symmetry. We show that non-linear quartic
mode-coupling terms allowed by symmetry enhances the Zeeman-field range over which the skyrmion-crystal phase remains stable. 
Furthermore, it leads to a significant dependence of the lattice constant of this spin crystal over this wide field range.
Conduction electrons coupled to such a tunable spin crystal are shown to experience a Berry-flux density which varies with the Zeeman field.
Such tunable skyrmion crystals provide a distinct realization of a Berry-Hofstadter butterfly, resulting in a phenomenon we term 
``topological quantum oscillations'' in the electronic density of states and associated observables.
\end{abstract}

\maketitle

Skyrmions were first introduced in particle physics as spatially localized topologically nontrivial field configurations to model nucleons. \cite{Skyrme1962}
In a condensed matter setting,
skyrmions appear as topological magnetic swirls in quantum Hall systems, \cite{Sondhi1993,Barrett1995,Shkolnikov2005} bulk and thin film
magnets, \cite{Rossler2006,BinzPRL2006,Muhlbauer2009,Tokura2010,Rosch2012,Tonomura2012,Seki2012}
and artificially engineered systems such as magnetic multilayers 
 \cite{Hoffmann2015, Kawakami2018} and oxide interfaces \cite{Banerjee2014,Xiaopeng2014,Matsuno2016,Mohit2016}.
Recent activity in this field has focused on the topological Hall effect in ferromagnetic skyrmion crystals, \cite{Ong2009,Neubauer2009,Kanazawa2011}
and on the hunt for skyrmions in frustrated antiferromagnets. \cite{Martin2008,Okubo2012,Leonov2015,Nagaosa2015, Mertig2017}

In many solid-state materials, such as MnSi, skyrmions occur as crystals --- periodic noncoplanar textures from a superposition of multiple
spin spirals \cite{Rossler2006,BinzPRL2006,Muhlbauer2009,Neubauer2009,Tonomura2012}
created by the Dzyaloshinskii-Moriya (DM) exchange interaction which is present in noncentrosymmetric systems.
An electron moving in such a spin texture
picks up a Berry phase. \cite{Volovik1987,Ye1999,Nagaosa2000}
The resulting Berry flux, arising from the scalar spin chirality of the skyrmion crystal, modifies the conventional Hall effect due to an orbital magnetic 
field and leads to an observed additional ``topological Hall effect'' contribution. \cite{Ong2009,Neubauer2009,Kanazawa2011}
Skyrmion crystals may also arise in frustrated antiferromagnets, and have been proposed to exhibit a 
larger scalar spin chirality \cite{Okubo2012,Leonov2015} and even a quantized and tunable
topological Hall effect.\cite{Martin2008,Okubo2012,Leonov2015,Nagaosa2015, Mertig2017,Pujol2019}

%\textcolor{red}{Skyrmion crystals may also arise in frustrated antiferromagnets}, and can exhibit a 
%large scalar spin chirality \cite{Okubo2012,Leonov2015} and even a quantized topological Hall effect.
%  \cite{Martin2008,Okubo2012,Leonov2015,Nagaosa2015, Mertig2017}
%Recent experiments \cite{kurumaji2018} have reported
%skyrmion crystals in a metallic layered-triangular antiferromagnet Gd$_2$PdSi$_3$. In general, such
%frustrated antiferromagnets can host stronger spin twisting and smaller skyrmions, permitting denser information storage. 

In this paper, we explore the
idea that if one can manipulate the skyrmion crystal, one can control the Berry flux associated with its nontrivial topology. If this can
be achieved with purely a Zeeman magnetic field, which couples to the local moments but 
not directly to the orbital motion of the electrons, it yields a new mechanism for quantum oscillations arising from an emergent magnetic field,
distinct from Landau quantization in an external orbital magnetic field. Exploring such novel mechanisms is of broad interest in
light of experiments which have discovered unexpected quantum oscillations in correlated Kondo insulators.
\cite{Sebastian_Science2015,Li_Science2018}

With this motivation in mind, we investigate a microscopic model for a chiral ferromagnet on the two-dimensional (2D) triangular lattice.
We show that a symmetry-allowed quartic anisotropy
permits strong nonlinear mode-coupling interactions between the spiral spin modes and the uniform magnetization 
mode.  In the presence of a 
Zeeman field, we find that this mode coupling tunes the spiral wavevectors, leading to 
skyrmion lattices which are highly field-tunable. For 3D cubic multiferroic Cu$_2$OSeO$_3$, \cite{Seki2012} a similar effect 
has been experimentally reported and discussed within a continuum Landau theory. \cite{Rosch2018}

Increasing the Zeeman field in this skyrmion crystal
leads to a 
decrease in the magnitude of the spiral wavevector. This
results in an expansion of the skyrmion lattice, leading to a decrease in the Berry-flux density.
A computation of the band structures of electrons 
Kondo coupled to this ``expandable'' spin texture reveals the formation of Chern bands. Indeed, we show that this
setup provides a realization of a tunable Berry-Hofstadter model driven by real-space Berry fluxes, offering
a new arena where such a tunable Hofstadter model \cite{Hofstadter1976} can be realized in experiments beyond
existing solid-state examples.
\cite{Weiss1990,Weiss2001,Schweizer2004,Cuniberti2007,Kim2013,Burgdorfer2014}
Due to the field tunability of the skyrmion lattice, we
show that even a pure Zeeman  magnetic field acting on the local moments leads to quantum oscillations in the electronic density of states. This
remarkable effect, which effectively arises from scanning through this novel realization of the Hofstadter butterfly spectrum,
should manifest itself in various physical observables, e.g., the electronic specific heat and transport of such skyrmion metals. We term this
phenomenon ``topological quantum oscillations''.

\section{Model} 

We begin with a  Hamiltonian describing  classical spins on a triangular lattice, which has the key ingredients previously shown \cite{Butenko2010,Banerjee2014,Xiaopeng2014,Mohit2016} 
to stabilize 2D skyrmion crystals within a Landau theory:
\bea
\!\!\!\! H_0 \!\!&=&\!\! - J_1 \sum_{\la i j \ra} \bS_i\cdot\bS_j + D \sum_{\nn} \left(\hatz \times \hat{r}_{ij}\right).\left(\bS_i\times\bS_j\right) \nonumber\\
\!\!\!\! &-&\!\! A_c \! \sum_{\nn} \left(\bS_i \! \cdot \! \left(\hat{z}\times \hat{r}_{ij} \right)\right)\left(\bS_{j} \! \cdot \! 
\left(\hat{z}\times \hat{r}_{ij} \right)\right) - {\cal B} \! \sum_i S_{i z}.
\label{eqH0}
\eea
We set the lattice constant of this underlying triangular lattice to be $a\!=\!1$. Here $J_1 \!>\! 0$ is the nearest-neighbor ferromagnetic Heisenberg exchange. Spin-orbit coupling induces anisotropic exchange interactions: $A_c$ is a compass term which is symmetry-allowed even in the presence
of inversion symmetry, while $D$ is the chiral Rashba DM interaction which arises only when $\hat{z} \to -\hat{z}$ mirror symmetry is broken. 
The microscopic derivation of this term in Rashba metals has been recently revisited.\cite{BrataasPRL2018a}
Dresselhaus DM interaction however is not allowed by vertical-plane reflections $\sigma_v$ of the triangular lattice.
The Zeeman field ${\cal B}$ allows for control of the ground state
of the system. Increasing ${\cal B}$ leads to a transition \cite{Banerjee2014,Mohit2016}  from a single-mode spiral order to a triple-{\bQ} skyrmion crystal, and eventually to a
polarized ferromagnet at large ${\cal B}$.
We supplement $H_0$ with two terms
\bea
H_1 &=& A_s \! \sum_i S_{i z}^2 + \lambda_\perp \sum_i S^4_{i z}
\label{eqH1}
\eea
Here $A_s$ is a single-ion anisotropy which has been previously shown to influence the stability of skyrmion-crystal order.\cite{Bogdanov2002,Banerjee2014,Mohit2016} 
The ingredient we focus on here
is the local anisotropy $\lambda_{\perp} > 0$ which is the single nontrivial {\it quartic} coupling allowed  by the $C_{6 v}$ symmetry of the lattice
for the case when $\bS_i^2 = 1$. As discussed
below, this term leads to 
a strong mode-mode coupling and field-tunable skyrmion lattices. Our results are obtained via an extensive variational optimization for the 
skyrmion crystal solutions discussed below.

\section{Variational phase diagrams} 

To study the ground states of the Hamiltonian $H_0 + H_1$, we resort to two different variational approaches: a momentum space approach and a 
real space approach. Within the momentum space approach, we use a Luttinger-Tisza method to identify the preferred wavevectors
$\bQ$ for spin-spiral solutions, which includes the uniform ferromagnetic mode as a special case with $\bQ=0$, and allows us to construct
skyrmion solutions as multi-$\bQ$ orders. Within the real space approach, we directly write down a periodic real space ansatz for various orders.
We discuss both methods below, and the resulting phase diagrams which share certain common features (which we consider reliable), and differ in
certain respects which we attribute to the drawback of the approximations.

\subsection{Momentum space approach}

In this approach, we begin by assuming that the local spin-length constraint $\bS_i^2=1$ is only satisfied on average,as $\sum_i \bS_i^2=N$ 
where $N$ is the number of sites. For $\lambda_\perp \! = \! {\cal B} \! = \! 0$, a Luttinger-Tisza analysis, which treats the spin Hamiltonian as 
a Gaussian field theory, imposing the spin length constraint only on average, yields degenerate spin-spiral solutions with certain
preferred wavevectors and their higher harmonics. We thus consider a general ansatz $\bS_i = {\bf m}_0 + \sum_n {\bf m}_n e^{i \bQ_n \cdot \br_i}$, retaining the uniform 
component ${\bf m}_0$, and a set of symmetry-equivalent spirals with wavevectors $\bQ_n$ and amplitudes ${\bf m}_n$.

The simplest skyrmion lattice is constructed from a set of three primary spiral 
wavevectors which sum to zero (in addition to the FM mode), i.e. it is a triple-$\bQ$ spin crystal.
The impact of the hard-spin constraint and the nonlinear
coupling $\lambda_\perp \neq 0$ is to generate higher harmonics \cite{BinzPRL2006} of these primary ${\bf Q}$'s.
The full variational state 
is obtained by optimizing the ansatz for $\{{\bf m}_0,{\bf m}_n,\bQ_n\}$ with several harmonics (see Appendix \ref{appendix:A} for details), 
followed by explicitly normalizing the spin vector at each site. We then compute the variational phase diagram by comparing the energies of the resulting different hard-spin
configurations. We note that our ansatz is general enough to encompass various phases: uniform ferromagnets, spirals, and multi-$Q$ skyrmion crystals.

\begin{figure}[t]
\makebox[0.1\linewidth]{\includegraphics[width=1.05\linewidth,keepaspectratio]{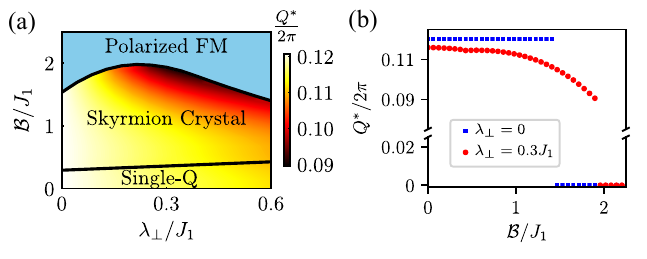}}
\caption{(a) Phase diagram of the triangular chiral ferromagnet as a function of nonlinearity $\lambda_\perp$ and magnetic field ${\cal B}$, showing single-$\bQ$ spiral, FM, and skyrmion crystal phases. We have fixed $D = J_1$, $A_c=0.4 J_1$ and $A_s=0.1 J_1$. Color scale shows optimal momentum $Q^*=|\bQ^*|$, featuring strong ${\cal B}$-dependence with increasing $\lambda_\perp$.
(b) Field dependence of $Q^*$, at $\lambda_\perp=0$ and $\lambda_\perp = 0.3 J_1$, displaying strong tunability of $Q^*$ over a wide field range 
in the presence of a nonzero $\lambda_{\perp}$ within the skyrmion-crystal phase.}
\label{fig:fig1}
\end{figure} 

{\bf Phase diagram:} The phase diagram obtained in this manner, as a function of ${\cal B}$ and $\lambda_\perp$, is depicted in 
Fig.~\ref{fig:fig1}(a).
For $\lambda_\perp=0$, this is consistent with previous studies,\cite{Banerjee2014,Mohit2016} exhibiting
spiral, skyrmion crystal, and FM phases with increasing ${\cal B}$. Our key observation is that incorporating $\lambda_\perp \neq 0$ leads to qualitatively similar phases. The key difference is that the skyrmion crystal window initially expands in the presence of $\lambda_\perp \neq 0$ before eventually shrinking. 
At the same time, the primary wavevectors $\bQ^*$, which lie in the
$\Gamma$-$M$ directions, now exhibit a smooth field dependence over a wide field range, as seen from Fig.~\ref{fig:fig1}(b).
In real space, this corresponds to a smooth Zeeman magnetic field
dependence of the skyrmion-crystal lattice constant over a wide range of fields.
We trace this dependence of $\bQ^*$ on ${\cal B}$ to an effective renormalization of the easy-plane
anisotropy. This is heuristically seen from a Hartree-level treatment \cite{Rosch2018} of the $\lambda_\perp$ quartic term, which yields
$A_s^{\rm eff} = A_s + 6 \lambda_\perp m_0^2$, where $m_0$ increases with increasing ${\cal B}$. This renormalization of $A_s$ leads
to a change in the optimal $\bQ^*$ over a wide range of fields, thus tuning the skyrmion crystal.

\subsection{Real space approach}

We next turn to a real space approach utilizing the well-known circular cell ansatz for skyrmion crystals \cite{Banerjee2014,Mohit2016} which is constructed so as to
directly obey the hard-spin constraint at each site. For large skyrmions, with $Q^*/2\pi \ll 1$, as in Fig.~\ref{fig:fig1}, it is useful to
treat the lattice spins $\bS_i$ as a smooth function of position  $\bS(\br)$. We can then obtain the energy functional for this spin field by Taylor expanding
the spins on bond $(i, i+\delta)$ as
\bea
\!\!\! \bS(i) \!\!&\approx &\!\! \bS(\br) - \frac{1}{2} \delta_{\alpha}\partial_{\alpha} \bS(\br) + \frac{1}{8}\delta_{\alpha} \delta_{\beta} \partial_{\alpha}\partial_{\beta} \bS(\br) \\
\!\!\! \bS(i+\delta) \!\!&\approx &\!\! \bS(\br) + \frac{1}{2} \delta_{\alpha}\partial_{\alpha} \bS(\br)
+ \frac{1}{8} \delta_{\alpha} \delta_{\beta} \partial_{\alpha} \partial_{\beta} \bS(\br)
\eea
where $\br \equiv (i+\delta/2)$. Inserting these into the Hamiltonian, we obtain the energy functional below
\beas
E &=& \frac{\sqrt{3}}{2} \int \td x\td y ~\mcE, \\
\mcE &=& \mcE_J + \mcE_D + \mcE_c + \mcE_a - \mathcal{B}S_z(\br),
\eeas
where the energy density $\mcE$ is a sum of various terms in the original Hamiltonian.
\beas
\mcE_J &=& \frac{3}{2}\frac{J_1}{2} \sum_{\alpha = x, y, z}(\nabla S_{\alpha})^2,\\
\mcE_D &=& \frac{3}{2}D\left[(S_z \partial_x S_x - S_x \partial_x S_z) + (S_z \partial_y S_y - S_y \partial_y S_z)\right],\\
\mcE_c &=& - \frac{3}{2} A_c (S_x^2 + S_y^2) \\
&& + \frac{3}{2}\frac{A_c}{8} \left[(\partial_x S_x + \partial_y S_y)^2 - 3 (\partial_x S_y - \partial_y S_x)^2\right],\\
\mcE_a &=& A_s S_z^2 + \lambda_{\perp} S_z^4.
\eeas
In arriving at these expressions, we have dropped overall constants and total derivative terms, and retained terms upto and including 
second order derivatives.
Following Refs.\onlinecite{Banerjee2014,Mohit2016}, we can then represent the various phases discussed within the momentum space approach 
by the following ansatze.

\noindent (1) {\bf Skyrmion crystal}: We consider a triangular skyrmion lattice, with spins inside each hexagonal unit cell being given by
 of the
\beas
\bS_{\text{skx}}(\br) &=& \left(\frac{x}{r} \sin \Theta(r), \frac{y}{r}\sin \Theta(r), \cos\Theta(r)\right),
\eeas
where the origin for this skyrmion is at defined to be at the center of the hexagonal unit cell, and
$\Theta(r)$ is optimized assuming it varies smoothly and radially within a disk embedded in the hexagonal unit cell. At the core,
$\Theta(0) = \pi$ and at the edge $\Theta(R_{\text{skx}}) = 0$. In the rest of the unit cell, $\Theta$ equals to 0.

\noindent (2) {\bf Spiral order}: For this phase, the spin field is assumed to spiral in the y-direction is given by a vector field with a periodicity 
$R_{\text{spi}}$ along the y-direction
\beas
\bS_{\text{spi}}(\br) &=& (0, \sin\theta(y), \cos\theta(y)),
\eeas
where $\theta(0) = 0$ and $\theta(R_{\text{spi}}) = -2 \pi$ for $D > 0$. We have chosen the spiral wavevector to be along y-direction based on the 
momentum space picture, within which $\bQ^*$ lies in the $\Gamma-M$ direction of the Brillouin zone. \footnote{In fact, the current form of $E$, limited to second order in
gradients, is invariant under any rotation around the z-axis, which can be seen by noting that $(S_x, S_y, S_z)$ transforms as $(x, y, z)$ under rotations. Thus the energy functional $E$ does not prefer any spiral directions.}

\noindent (3) {\bf Uniform FM}: The ansatz for the FM allows for a canting angle $\varphi$ with respect to the field direction,
so that the the spin field is given by
\beas
\bS_{FM} &=& (\sin \varphi, 0, \cos\varphi).
\eeas

\begin{figure}[t]
\centering
\includegraphics[width=\linewidth]{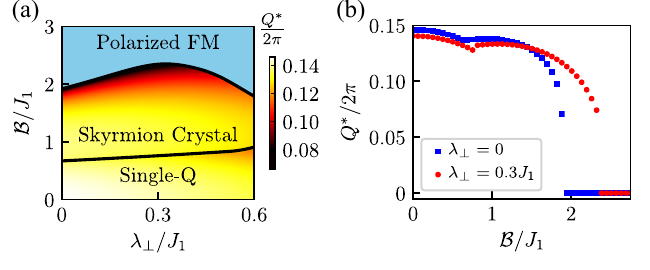}
\caption{(a) Ground state phase diagram as a function of the Zeeman field $\mathcal{B}$ and the nonlinearity coupling $\lambda_{\perp}$ for the continuum energy functional discussed in the main text. The color gradient shows how $\lambda_{\perp}$ leads to a field-tuned periodicity of the skyrmion crystal. (b) Field dependence of $Q^*$ at $\lambda_{\perp} = 0$
and $\lambda_\perp = 0.3 J_1$. For $\lambda_\perp=0.3 J_1$, this features an extended field range with a skyrmion crystal phase having a tunable $Q^*$. This is in agreement with the multi-Q variational phase diagram for the lattice model in the main text. When $\lambda_\perp=0$, then $Q^*$ only shows rapid variation in the close vicinity of the skyrmion-crystal-to-FM phase boundary, providing a very narrow window where the emergent topological flux density has significant variation.}
\label{fig:fig2}
\end{figure}

{\bf Phase diagram:}
We carry out a functional minimization of $E$ with respect to $\{R_{\text{skx}}, \Theta(r)\}$, $\{R_{\text{spi}}, \theta(y)\}$ and $\varphi$ in the different ansatze, and compare
energies to arrive at the phase diagram in Fig.\ref{fig:fig2}(a) which qualitatively resembles the momentum space phase diagram, showing spiral, skyrmion crystal, and uniform
polarized FM phases. We see that the nonlinear coupling $\lambda$ enhances the skyrmion crystal window, as in the momentum space approach.
The color scale displays the optimal wavevector $Q^*$,
defined by $2\pi/R^*_{\text{spi}}$ in the spiral phase, and as $2\pi/\sqrt{3}R^*_{\text{skx}}$ in the skyrmion crystal phase, where $R^*_{\text{spi(skx)}}$ is the optimal value of 
$R_{\text{spi(skx)}}$, and Fig.~\ref{fig:fig2} shows a cut through the phase diagram displaying the change of $Q^*$ with the Zeeman field for zero and nonzero $\lambda$.

\subsection{Tunable skyrmion crystal}

We next compare the momentum space and real space approaches to identify the common features which we consider to be reliable. We also highlight 
differences which we attribute to the drawbacks of the different
approximations.

At a broad level, both approaches lead to similar phase diagrams, showing regimes of spiral, skyrmion crystal, and polarized FM orders. In both approaches,
the additional anisotropic nonlinearity $\lambda$ is seen to initially enhances the regime of the skyrmion crystal at the expense of the polarized FM, while the skyrmion
crystal shrinks at large $\lambda$. We find that the real space approach which strictly respects the spin length constraint, leads to a slightly larger regime of stability of the
spiral order relative to the skyrmion crystal when compared with the momentum space approach. 

The field dependence of the skyrmion crystal wavevector
is somewhat different in the two approaches. Our momentum space Luttinger-Tisza based ansatz at $\lambda=0$, reduces to a harmonic problem, which clearly does not
allow for $Q^*$ to depend on ${\cal B}$, except that it jumps to $Q^*=0$ when the skyrmion crystal gives way to the polarized FM. This is a drawback of our momentum space
approach. The real space approach by contrast is inherently
nonlinear, and permits the wavevector to change somewhat with ${\cal B}$, except that the most dramatic variation is a rapid drop near the skyrmion crystal to FM 
phase boundary (rather than a jump). We expect this real-space result for $Q^*$ to be more reliable when $\lambda \!=\! 0$.

However, in both approaches, having $\lambda\neq 0$ leads to a smooth variation of $Q^*$ over a very wide ${\cal B}$-field regime in the skyrmion crystal phase. 
We thus conclude that the generically allowed nonlinearity $\lambda \neq 0$ leads to robust feature of a Zeeman field tunable spin crystal. The impact of this
field-tunable skyrmion crystal on electronic response will be discussed below.

\begin{figure}[b]
\includegraphics[width=0.49\textwidth,keepaspectratio]{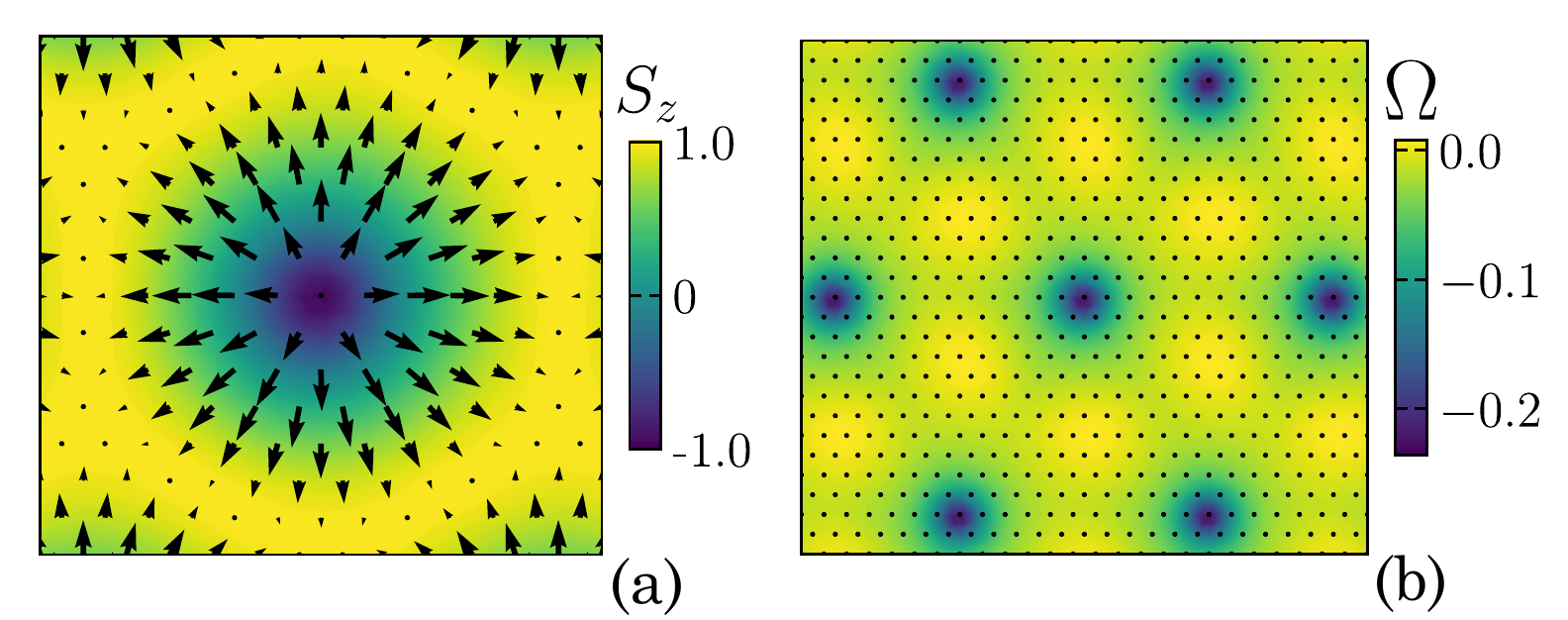}
\caption{(a) Structure of N\'eel skyrmion with negative $S_z$ magnetization at the core, and in-plane spin component pointing radially
outward from the core. (b) Spatial variation of solid angle $\Omega$
subtended by noncoplanar spins on elementary triangular plaquettes. Normalized solid angle
$\Omega/4\pi$ summed over the lattice yields a topological charge $(-1)$ per skyrmion.
\label{fig:texture}
}
\end{figure}

\subsection{Plot of skyrmion texture}

The spin texture stabilized in both approaches is a lattice of N\'eel skyrmions, as shown in Fig.~\ref{fig:texture}(a) from the projected
momentum space ansatz. We find that the solid angle $\Omega$
subtended by the three spins on each elementary triangular plaquette shows a strong spatial variation,
being higher at the skyrmion core, as
seen from Fig.~\ref{fig:texture}(b). (In the continuum, $\Omega = \bS\cdot\partial_x\bS\times\partial_y\bS$.)
Adding up $\Omega/4\pi$ across the lattice, we obtain 
an average topological charge $(-1)$ per skyrmion.
When the Zeeman field enhances the skyrmion-crystal lattice constant, this flux gets distributed over a larger area. We 
thus expect electrons moving in such
a spin-crystal background to sense an average ``emergent orbital field'' which decreases as we increase the Zeeman field on 
the local moments.

\section{Topological quantum oscillations} 

We finally turn to the nature of electronic bands which emerge from coupling conduction electrons to highly tunable skyrmion lattices.\footnote{Within
the momentum space approach, we have directly constructed the skyrmion crystal on the lattice, while in the real space approach we have
taken the continuum limit appropriate for large skyrmions as studied here. In the latter case, we much thus
place the spin field onto the underlying triangular lattice in order to couple it to electrons.} 
We consider a nearest-neighbor ordinary hopping $t$ and Rashba hopping $\chi_R$ for the conduction electrons, and
assume that the electrons sense \cite{Nagaosa2000, Nagaosa2015} the underlying spin texture $\{ \bS_i\}$ via a ferromagnetic Kondo interaction
$J_H > 0$  which can physically arise from atomic Hund's coupling.
To study the influence of the skyrmion crystal on conduction electrons, we have diagonalized a 
Hamiltonian for electrons on the triangular lattice (same lattice as the spins),
\bea
H_{\rm elec} &=& - \! \sum_{j,\ell} t^\pdg_{j\ell} c^\dg_{j \alpha} c^\pdg_{\ell \alpha} \! - \! \sum_{j,\ell} i \chi^{(j\ell)}_R c^\dg_{j\alpha} \left[ \hat{z} \times \hat{r}_{j\ell} 
\! \cdot \! \vec\sigma \right]_{\alpha\beta} c^\pdg_{\ell\beta} \nonumber \\
&-& J_H \sum_j \bS^\pdg_j \cdot c^\dg_{j\alpha} \vec\sigma^\pdg_{\alpha\beta} c^\pdg_{j\beta}.
\eea
For a ``magnetic unit cell'' of $n^2$ sites, we diagonalize the corresponding $2 n^2 \times 2 n^2$ electronic 
Hamiltonian at all momenta $\bk$ in the
reduced hexagonal Brillouin zone (BZ) of the skyrmion crystal. 
The resulting number of electronic bands depends on the skyrmion lattice.
If ${\bQ^*} = (4\pi/\sqrt{3}) (m/n) \hat{y}$ (with its symmetry-allowed partners) 
is the best approximant to the optimal spin-crystal wavevector, 
the periodic magnetic unit cell of the skyrmion crystal has $n^2$ triangular lattice sites and encloses $m^2$ skyrmions.
This hosts $2n^2$ bands including spin.
Within the real space approach to the skyrmion crystal, we only permit $m=1$.
 In our work, we considered magnetic
unit cells of up to $n^2 =3600$ sites  (i.e., $n \leq 60$) with varying ${\cal B}$.
To compute the Chern number of the electronic bands, we have used
the method proposed by Fukui, Hatsugai, and Suzuki \cite{Fukui2005} on a finely discretized momentum mesh in the BZ.

\begin{figure}[tb]
\centering
\includegraphics[width=0.48\textwidth,keepaspectratio]{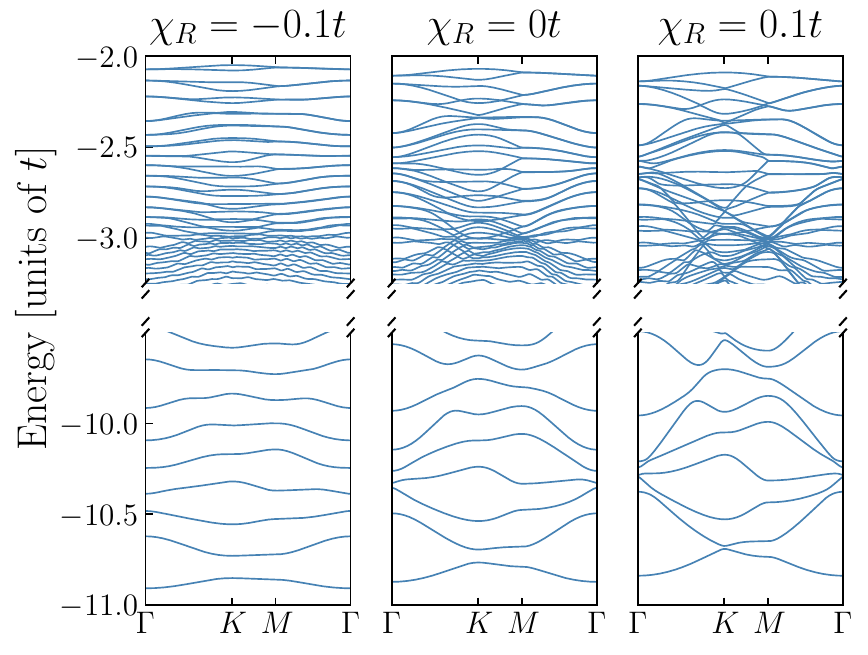}
\caption{Band structure along high-symmetry paths in the hexagonal BZ 
in the skyrmion-crystal phase for the ferromagnet with DM interactions. The conduction electrons are assumed to
have ordinary hopping $t$, a Rashba hopping $\chi_R=0,\pm 0.1t$, with a coupling to the localized spins via ferromagnetic
exchange $J_H/t=5$. The plotted bands correspond to half of all the bands, where conduction-electron spins are aligned 
with the local moments. The low-energy bands are well isolated, and carry Chern numbers $-1$. The high-energy bands
come in pairs, exhibiting band sticking along $\Gamma$-$M$, with total Chern number $C_{\rm pair}= -2$. Finally, the
set of bands near $E \approx -3t$ which are connected and carry large $C>0$ Chern numbers ensure that the
total Chern number is zero.
\label{fig:bandrashba}
}
\end{figure}

This commensurability effect of the skyrmion lattice with the
underlying crystal lattice does not appear in continuum field theories. It has also not been discussed in previous work
on lattice models, \cite{Nagaosa2015, Mertig2017} since these have only studied the simplest example with fixed $m=1$.
We find that in the case with $m > 1$, the Berry flux exhibits small 
modulations from one skyrmion to the next, with periodicity recovered only over the magnetic unit cell which encloses $m^2$
skyrmions. Crucially, we find that adding up the solid angles subtended by the
three spins on an elementary plaquette leads to a
quantized topological charge $(-1)$ per skyrmion only when we average over {\it all} $m^2$ skyrmions in the magnetic unit cell.

An example of the resulting electronic band dispersions for $\chi_R/t = 0,\pm 0.1$ and $J_H/t = 5$ are shown
along high-symmetry paths in the BZ in Fig.~\ref{fig:bandrashba}. The plotted bands correspond to half of all the $2 n^2$ bands, where conduction-electron spins are aligned 
with the local moments. In all cases, these low-energy bands appear isolated and have 
Chern number $C=-1$, resembling somewhat dispersive Landau levels. Thus, at special electronic densities which fill up an integer number of these bands,
we would expect a quantized topological Hall effect. \cite{Nagaosa2015, Mertig2017}
The high-energy bands come in pairs due to
band sticking along the high-symmetry $\Gamma$-$M$ line; in this case the total Chern number of the pair is $C_{\rm pair}=-2$.
In an intermediate-energy window, around $E/t \! \approx \! -3$, we find a set of bands carrying large positive Chern numbers $C \gg 1$,
which ensures that the Chern numbers over all bands sum to zero.
For $|\chi_R| \ll t$, this occurs around the van-Hove singularity in the density of states (DOS) of the triangular lattice,
where a Lifshitz transition occurs near the $M$-point. This distribution of Chern numbers is analogous to that for Chern bands in the
Hofstadter model. \cite{Hofstadter1976,Mertig2017}

\begin{figure}[t]
\centering
\includegraphics[width=0.48\textwidth,keepaspectratio]{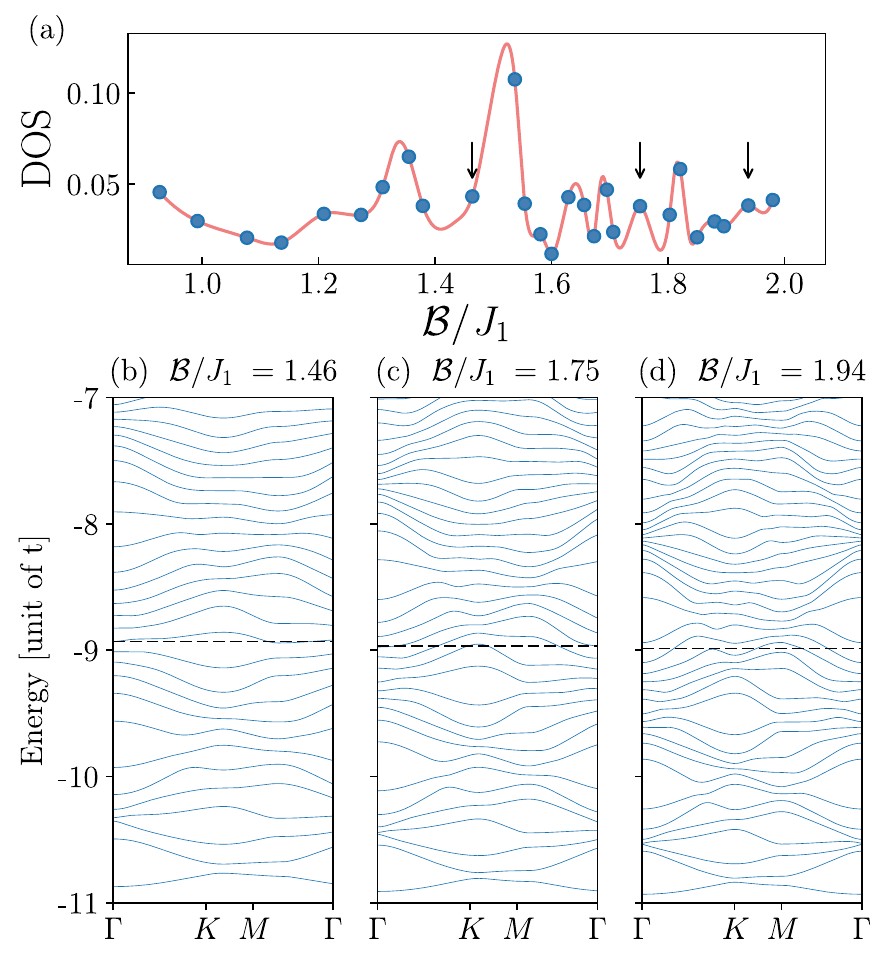}
\caption{(a) Oscillations
in the DOS at the Fermi level (in units of $t^{-1}$) for $\chi_R=0$ and fixed electron density of $\rho=0.1$ electrons per site. 
These topological quantum oscillations arise from the ${\cal B}$ dependence of $Q^*$. Solid line is a guide-to-the-eye computed by considering 
small variations of the density around $\rho=0.1$. Arrows indicate the positions in $\mathcal{B}$ space, whose band structures are shown in (b)-(d). They correspond to textures with optimal momenta $\bQ^*=(4\pi/\sqrt{3})(m/n)$ with $m = 1$ and $n = 11, 12, \text{ and } 13$ respectively. The dotted lines denote the chemical potentials, which appear to be fixed, while the number of occupied Chern bands changes from approximately (b) 12 to (c) 14 to (d) 17.}
\label{fig:dos}
\end{figure}

Fig.~\ref{fig:dos}(a) plots the electronic DOS at the Fermi level as we vary the Zeeman field ${\cal B}$ 
on the local moments, staying at a fixed electron density $\rho=0.1$ electrons per site of the original triangular lattice. 
The textures are the variational hard-spin configurations obtained from the multi-$Q$ calculation 
at $\lambda_{\perp} = 0.3 J_1$ (see Fig.\ref{fig:fig1}(b).) Despite the low energy
Chern bands having some dispersion, we find that the DOS shows clear oscillations, even though the electrons are not 
subject to any orbital magnetic field. In analogy with the terminology ``topological Hall effect'', we term these as
``topological quantum oscillations''. 

Fig.\ref{fig:dos}(b)-(d) show how band structures vary with the Zeeman field via its impact on the spin texture. 
This is different from the previous studies\cite{Nagaosa2015, Mertig2017} 
where electronic properties are examined by simply tuning the chemical potential in a fixed spin texture. In our case, the
$Q^*$ values at the indicated Zeeman fields correspond to ${\bQ^*} = (4\pi/\sqrt{3}) (m/n) \hat{y}$ with 
$m = 1$ and $n = 11, 12, \text{ and } 13$. For the density $\rho = 0.1$, the number of the occupied Chern bands varies from approximately $12$ to $14$ to $17$.

The oscillations can be understood by analogy with the Hofstadter model. When conduction electrons move in a closed loop aligning their spins
with an underlying noncoplanar texture,they pick up a Berry phase which is analogous to an Aharonov-Bohm phase.
\cite{Volovik1987,Ye1999,Nagaosa2000,Trivedi_PRL2012}  As a result, each spin texture configuration may be viewed as an orbital flux configuration for conduction electrons,
resulting in a Hofstadter-type model.
When the real space Berry curvature is not changing, it corresponds to a constant fictitious orbital magnetic field. When the Zeeman field causes an
expansion of the skyrmion lattice constant, it thus leads to a decrease in the average Berry-flux density. This problem is
thus analogous to studying a Hofstadter model with a {\it decreasing} orbital magnetic flux $\phi$ per plaquette as we increase the Zeeman field ${\cal B}$.
Thus, increasing the Zeeman field $B$ in our case leads to {\it more} ``Landau-like levels'' below the Fermi energy as is clear from Fig.~\ref{fig:dos}(b)-(d).

In the conventional Hofstadter problem, \cite{Hofstadter1976,Weiss1990,Weiss2001,Schweizer2004,Cuniberti2007,Kim2013,Burgdorfer2014}
varying $\phi$ leads to a rich
Hofstadter butterfly spectrum due to
the interplay of two length scales --- set by the unit cell area of the underlying lattice and the area of the magnetic unit cell.
In our problem, the role of $\phi$ is effectively played by $Q^{*2}= [(4 \pi/\sqrt{3})\  (m/n)]^2$.
As the Zeeman field $\mathcal{B}$ is tuned, $Q^*$ changes, and the oscillations in the DOS can thus
be identified with a scan in $\phi$ through the Hofstadter butterfly spectrum.
Our model differs from the conventional Hofstadter model in that the flux per plaquette is not 
really uniform. The flux inhomogeneity is characterized by 
a third length scale -- the skyrmion size. Most of the Berry flux attached to each skyrmion is concentrated 
in this region around the core, so this serves to characterize the spatial inhomogeneity of the emergent ``orbital magnetic field''. The
flux is most smeared out (i.e., the most homogeneous) when the skyrmion size is similar to the inter-skyrmion separation. 
However, this flux inhomogeneity only leads to quantitative changes, and does not qualitatively impact the results.

These topological DOS oscillations should impact all physical observables.
Since we are tuning the Zeeman field ${\cal B}$ on the local moments, this nontrivial effect is distinct from previous studies which 
simply varied the electron filling in a fixed skyrmion texture.
The oscillations we describe would be further impacted by the direct Zeeman and orbital effects
of the magnetic field on the conduction electrons; however, the typically much stronger effect of the skyrmion lattice and its
Berry flux should still lead to observable effects as deviations from conventional quantum oscillations in skyrmion magnets.

\section{Summary and discussion}
In summary, we have studied magnetic skyrmions in 2D chiral ferromagnets, arguing that quartic anisotropies can lead to field-tunable
skyrmion crystals via nonlinear mode-coupling effects. Such anisotropies can arise from single-ion physics or from integrating out conduction electrons.
For the case where this field tunability is highly pronounced, we have shown that topological quantum oscillations can arise from just the impact
of a Zeeman field on the local moments. The Chern bands which contribute to these oscillations are already somewhat dispersive; we thus expect
disorder broadening which is weaker than the scale of this bandwidth (which is also comparable to the band gap) will not alter our main results. The
amplitude of these topological oscillations are not expected to follow simple Lifshitz-Kosevich scaling since, unlike Landau levels, the emergent Chern bands are not
flat and equispaced in energy. The physics underlying these topological quantum oscillations is not crucially dependent on the hexagonal
symmetry of the underlying microscopic lattice, and thus
a heterostructure with an ultrathin metallic film deposited on insulating Cu$_2$OSeO$_3$, where a field-tunable skyrmion crystal has been reported,
\cite{Rosch2018} may be a potential venue for observing this effect. 
Further experimental and theoretical work on such a proposed heterostructure would be valuable in assessing the viability
of our proposal, as would searches for more promising material candidates.

\noindent We acknowledge useful discussions with Achim Rosch. 
This research was funded by NSERC of Canada and the Canadian Institute for Advanced Research. Computations were performed on the Niagara supercomputer at the SciNet HPC Consortium. SciNet is funded by: the Canada Foundation for Innovation under the auspices of Compute Canada; the Government of Ontario; Ontario Research Fund - Research Excellence; and the University of Toronto.

\appendix

{\section{Multiple-Q variational method}
\label{appendix:A}

\begin{figure}[b]
\center
\includegraphics[width=0.45\linewidth]{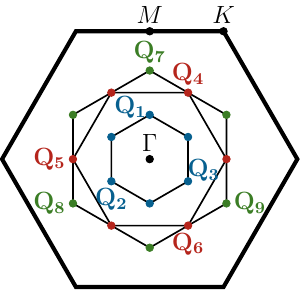}
\caption{Nine distinct wavevectors we have retained in the multi-$Q$ variational method. The primary wavevectors are $\bQ_1, \bQ_2, \text{ and } \bQ_3$, which are
related to each other by $C_3$ rotations. The higher harmonics modify the details of the texture without affecting the periodicity imposed by the primary $\bQ$'s. 
The wavevectors are, for instance, given by $\bQ_4 = \bQ_1 - \bQ_2$ and $\bQ_7 = 2\bQ_1$. The outer hexagon represents the Brillouin zone of the underlying 
triangular lattice.}
\label{fig:harmonic}
\end{figure}

The ground state energy of the spin Hamiltonian is minimized, assuming that the spin texture takes a multiple-Q superposition form
\bea
\bS_i = \bm_0 + \sum_{n=1}^9 \left(\bm_n \te^{\tI\bQ_n \cdot\br_i} + \text{c.c.}\right),
\label{eq:ansatz}
\eea
where $\bm_0$ and $\bm_n$'s are respectively real-valued and complex-valued three-dimensional vectors. Each harmonic is associated with a wavevector $\bQ_n$, and we have truncated keeping nine wavevectors (see Fig.\ref{fig:harmonic}). We implement the spin length constraint $\bS_i^2=1$ on average, namely $\sum_i \bS_i^2 = N_{\text{site}}$, 
which amounts to $|\bm_0|^2+ 2 \sum_n |\bm_n|^2 = 1$. 
The energy of the ansatz is minimized with respect to $\{\bm_0, \bm_n, \bQ_n\}$.
We then project this configuration to
real space and impose the strict hard-spin constraint to obtain the variational ansatz. 
For the spiral order, termed `single-$Q$', the variational ansatz may be further simplified using a restricted set of harmonics of Eq.\ref{eq:ansatz}, 
e.g. $\{\bQ_1, \bQ_7\}$ in Fig.\ref{fig:harmonic}, while for the spin-polarized FM, it is sufficient to retain only $\bm_0$.
The energies computed using these hard-spin variational configurations are used to 
determine the phase diagram.

\bibliography{references}

\onecolumngrid

\end{document}